# Users Approach on Providing Feedback for Smart Home Devices – Phase II

Santhosh, Pogaku

Student – MSIT School of Information Technology, University of Cincinnati, Pogakush @mail.uc.edu

**ABSTRACT**

Smart Home technology has accomplished extraordinary interest in making individuals' lives more straightforward and more relaxing as of late. Technology as of late brought about delivering numerous savvy and refined frameworks which advanced clever living innovation. In this paper, we will be investigating the behavioural intention of user's approach on providing feedback for smart home devices. We will be conducting an online survey for sample of three to five students selected by simple random sampling to study the user's motto for giving feedback on smart home devices and their expectations. We have observed that most users are ready to share their feedback on smart home devices actively to improvise the service and quality of the product to fulfill the user needs and make their lives easier.

**INTRODUCTION**

The current market of smart homes projects a global revenue of 85 Billion USD, and the usage is read to be around 10% worldwide[1]. Smart home technology combines automation interface, monitors, and sensors [2]. Smart home devices will tend to increase the way of life of humanity and the comfort of many people. The looks are deceiving as smart homes aren't taking the user needs and necessities into accountability [3] as several critical challenges related to innovative home technology are yet to be addressed for smart home development [4]. Smart home technologies are not very often in the way they should be used, even with specific settings. A clear view of the user's usage of the smart home devices is still missing. User feedback plays a vital role in getting the desired outcome and helps understand the user needs and what they want to use the smart home devices in the preferred way to fill the usage gap [4]. The essential features of smart home devices such as UI, Networking, and Interface push users to revisit or re-think the existing feedback approaches. For instance, from the feedback sender point of view, we have no idea to what extent the users who are using the smart home technology are genuinely providing feedback to the devices during their daily routines regarding their personal space as it's a sensitive area. From the feedback receiver's point of view, we have no clue what information the receiver needs from the feedback provided to improve his products. This paper performs a study on the users' behavioural intention on providing feedback on smart home devices.

Innovative home solutions are interconnected all the time, and the ecosystem has different types of devices, and a few of them are sensors where no human intervention is needed. Smart home products are other than regular products. These products are present in the living space of humankind. Yet, these devices even take care of us, and it brings a brand-new facet to such kind of feedback. Hence, from the industry and researchers' perspective, it is evident that there is a need for research to build excellent feedback tools or models[5]. The long-term research goal would be to construct group-based feedback solutions for smart devices. At the beginning of the design, the software and the engineering requirements will define the user feedback [6]. The feedback designed is significant and developed based on the existing experiences and would be helpful for future software development [7]. In the past, research has been conducted, and the toll support user feedback has been used, but it has dedicated settings[8] [9]. Many studies have projected that the behaviour of the feedback and software and preferences will vary accordingly, and it causes a mismatch.

This research investigates the user approach to providing feedback for smart home devices. The research question is as follows:
RQ: Behavioural intention of users to give feedback on smart home devices and contribute to making the devices better and user-friendly?

**METHODOLOGY**

In this study, we used quantitative survey methodology by randomly selecting students from the group assigned by Dr. Hazem Said to know the approach towards providing feedback on smart home devices. Using a simple random sampling method, we have selected a sample of the top 4 students representing 25% of the research participants group 2.

A survey instrument was used in this quantitative research approach. A structured questionnaire with four measures which is attached in "Appendix A," was designed utilizing the Behavior intention (BI) model [10] to measure the Behavioral Intention of user's approach on providing feedback for smart home devices, these measures were scored using a five-point where 1 = strongly disagree to 5 = strongly agree, and the measures were modified to match the technology used in the study. The adapted instrument's construct validity, internal consistency, and reliability of the scales were evaluated, validated, and proven.

The survey instrument was designed on the google forms platform, the survey questions were placed in a sequence, and the linear scale from 1 to 5 was added as an answering option having a description of 1 = strongly disagree to 5 = strongly agree. The questionnaire instrument was sent to experts to perform a field study to confirm its content validity. Based on the suggestions given by the experts, a few changes have been made, such as adding a description in the title for better understanding.

We sent a formal email via UC Canvas to the sample containing the survey URL on 03/08/2022, mentioned that the survey form is anonymous and that no personal data will be gathered, and urged the users to fill in the survey with genuine feedback. The sample was given two reminders on 03/09/2022 and 03/10/2022 via emails to fill in the survey. The survey was closed on 03/10/2022, and the four responses were saved in the google forms and later downloaded and used to conduct the in-depth analysis on the findings.

**RESULTS**

The data from the survey was collected by downloading an excel sheet from the options given in google forms. This data consists of all the responses given by the sample. We have rechecked any inconsistent information or incorrect data entered in the survey form to eliminate the possible errors. There is less possibility of error as we have implemented a field test. We have renamed the samples from sample-1 to sample- 5 to perform the statistical analysis. A column has been generated for the behavioral intentional questionnaire and named the column header a BI Questions and followed horizontally from BQ1 to BQ5. The sum of all four values for a given sample becomes a quantitative number representing the entire behavioural intention of one sample, this has been performed for all five samples, and each sample will get one quantitative number. Table 1 illustrates the projection of the analysis made to the data, and Figure 1 shows the responses given by the samples in a graphical representation.

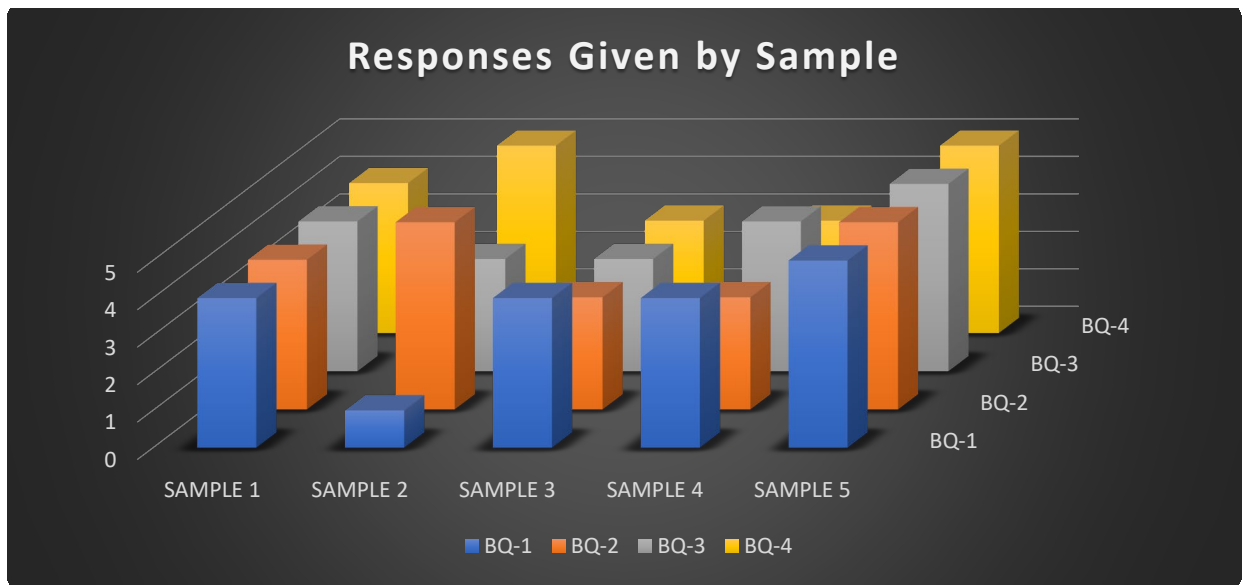

Figure 1: Responses Given by Sample

| BI Questions | SAMPLE 1 | SAMPLE 2 | SAMPLE 3 | SAMPLE 4 | SAMPLE 5 |
|---|---|---|---|---|---|
| BQ-1 | 4 | 1 | 4 | 4 | 5 |
| BQ-2 | 4 | 5 | 3 | 3 | 5 |
| BQ-3 | 4 | 3 | 3 | 4 | 5 |
| BQ-4 | 4 | 5 | 3 | 3 | 5 |
| SUM | 16 | 14 | 13 | 14 | 20 |

Table 1: Calculations made on the sample

| Question Tag | Questions |
|---|---|
| BQ-1 | I Intend to provide feedback for the smart home devices continuously? |
| BQ-2 | I will strongly recommend providing feedback that will help to improve the performance and service of the smart home devices. |
| BQ-3 | I plan to continue providing feedback frequently on the smart home devices and contribute to making the device better |
| BQ-4 | Assuming that I have very limited time, I intend to provide feedback frequently on the smart home devices |

Table 2: Representation of the Questions

After summing each sample, we observed two quantitative numbers repeating in the sample data, and 14 can be considered the mode. This suggests that there is repetitive behaviour in the sample. We have calculated the mean by summing up all the sample values, which gave us the value as 77 and divided the value by the no of samples five, and it resulted in ending up giving the mean value as 15.4. The derived mean value is critical to calculating the standard deviation and the standard error. We have calculated the standard deviation by using the formulae in the excel sheet as =STDEV("sampling range") and provided the values as (16, 14, 13, 14, 20), and the standard deviation was derived as 2.792. The standard error has been calculated by using the formulae = STDEV(sampling range)/SQRT(COUNT(sampling range)), and the result obtained as 1.248. We also calculated the variance using the formulae =var("sampling range"), and we got the variance value for the sample as 7.8.

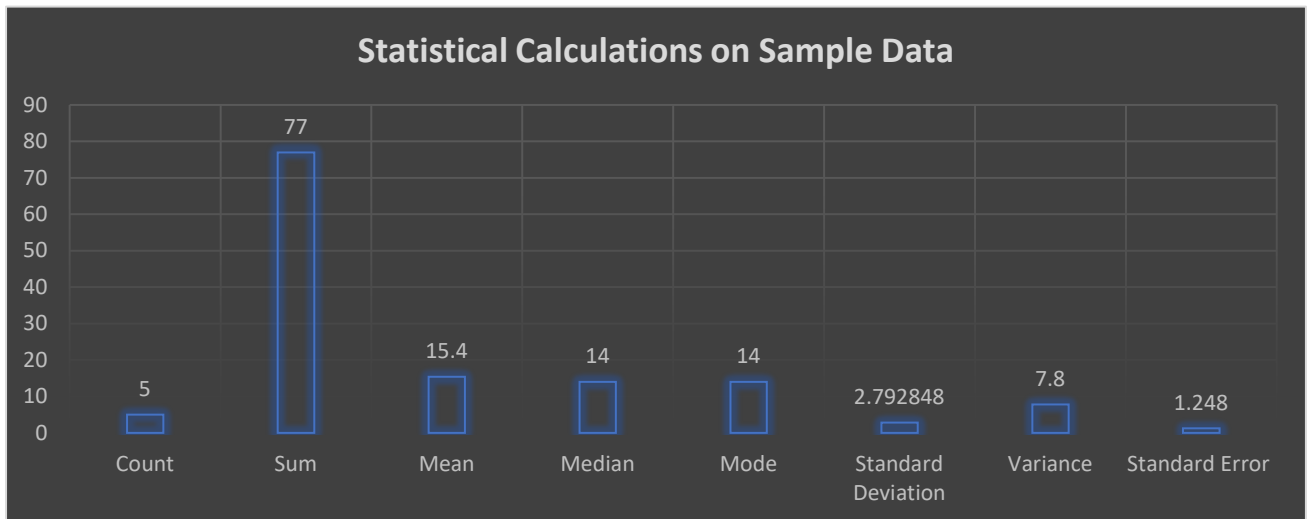

Figure 2: Representation of statistical Calculations.

Figure 2 represents the mean is 15.4, and the mean is greater than the sum of the BQs of samples 2, 3 and 4 and less than the sum of BQs of samples 1 and 5. This suggests that around 60% of the sample are inclined not to provide feedback for smart home devices, and 30% of the users are inclined to give feedback on the smart home devices to make them better. Figure 3 represents the standard deviation of the behavioral intentional model of the user's feedback approach on smart home devices. Three samples come under the 1st standard deviation, and only one comes under the 2nd standard deviation.

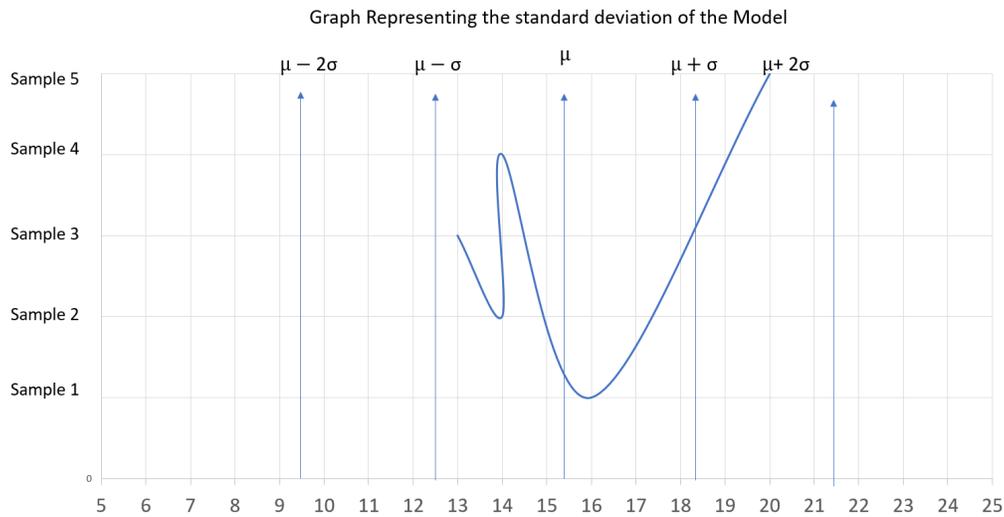

**Figure 3: Representation of statistical Calculations.**

**DISCUSSION**

This quantitative survey analysis shows how the users are interested in providing feedback on smart home devices and how the feedback can be used to improve the devices. This study was conducted via surveys using an online platform to know the user feedback approach on smart home devices. The sample's quantitative answers helped us understand how eager the users are to provide feedback to improve the machines and performance. We talked with understudies to a higher view on the conduct, mentality, and requirements for criticism arrangement on brilliant home gadgets. We accept that the different points of interaction for controlling apparatuses and devices and the attributes of savvy home innovations like network, universality, and even intangibility will be tested while practically applying a criticism approach.

There is a certain limitation to the data, such as having access to limited people in the sample frame, which may not meet the external validity as we were not having access to the larger sample and were limited to a certain sample. The quantitative investigation methodology incorporates a study with close-completed questions. It prompts limited outcomes shown in the investigation recommendation. In this way, the results can't constantly address the genuine occurring in a summarized structure. Moreover, the respondents have limited responses, considering the decision made by the investigator. Quantitative exploration requires significant measurable examination, which can be hard for non-analysts to do. Non-mathematicians might observe measurable examination testing since it relies upon logical discipline. We have selected around the sample size of 75% of students among the group, which directly impacts the transferability. The random selection of 12 students among 16 students shows the high dependability of the data we received from the students. The validity is established by taking the answers and providing graphs from the data presented in the results section. We have selected around the sample size of 25% of students among the group, which directly impacts the transferability. The random selection of 5 students among 16 students shows the high dependability of the data we received from the students. The validity is demonstrated by analyzing the responses and providing graphs from the data presented in the results section. We have conducted semi-structured interviews in the previous paper using qualitative measures to gauge the user's feedback on smart home devices and projected the results, and as part of phase 2, we have used survey methodology using quantitive measures and performed analysis.

The results align with the findings [2] C.Wilson and Tom Hargreaves, where the results of peer review research on smart home users and the exponential growth of the users. The author projected the functional view, instrumental view of the smart home devices. It also explained user-related challenges related to smart home challenges and lack of feedback where it matches how users are eager to provide feedback to improvise the smart home devices. After conducting a detailed survey, we have taken the quantitative data provided by the sample on a linear scale and performed statistical analysis to check the behavioural intention of the sample on providing feedback on smart home devices. Most of the sample answers were inclined to provide feedback, and around three samples' responses for two measures were neutral. After performing the statistical analysis overall, the study showed that the sample had a neutral behavioural intention to provide feedback on smart home devices.

We recommend performing even more in-depth studies by exploring and conducting surveys on the larger sample on providing feedback and how effective the feedback is given will help the sample and the receivers to change the design, performance, service, and usability of the smart home devices.

**REFRENCES**

# Appendix A

IT7010 Information Technology Research Methods

Survey Script

Hello,

Good evening. I hope you're doing well.

My self-santhosh Pogaku. I am part of the research team which conducts a survey on the Users Approach to provide Feedback for Smart Home Devices.

This research is anonymous and secure as we don't store any usernames or email on this form.

Survey Questions

1. I Intend to provide feedback for the smart home devices continuously?
2. I will strongly recommend providing feedback that will help to improve the performance and service of the smart home devices.
3. I plan to continue providing feedback frequently on the smart home devices and contribute to making the device better
4. Assuming that I have very limited time, I intend to provide feedback frequently on the smart home devices

Thanks for taking the time and participating in the survey.